\documentclass[english,12pt]{article}
\usepackage{lmodern}

\usepackage[T1]{fontenc}
\usepackage[latin9]{inputenc}
\usepackage{mathrsfs}
\usepackage{amsmath}
\usepackage{amssymb}

\makeatletter

\newcommand{\lyxaddress}[1]{
	\par {\raggedright #1
	\vspace{1.4em}
	\noindent\par}
}

\usepackage{commands}

\usepackage{cancel}
\usepackage{mathrsfs}   
\usepackage{slashed}     
\usepackage{bbold}  
\usepackage{url}
\usepackage{graphicx}
\usepackage[colorlinks=true,linkcolor=redLinks,citecolor=greenLinks,urlcolor=redLinks, pdfborder={0 0 1}]{hyperref}
\usepackage{xcolor}
\usepackage{framed}
\usepackage[numbers,sort&compress]{natbib}
\usepackage{amsmath}
\usepackage{microtype}
\usepackage[normalem]{ulem}
\usepackage{siunitx}

\usepackage{titlesec}
\usepackage{kbordermatrix}

\allowdisplaybreaks

\colorlet{shadecolor}{gray!15}

\definecolor{greenLinks}{rgb}{0, 0.6, 0} 
\definecolor{blueLinks}{rgb}{0, 0, 0.6}
\definecolor{redLinks}{rgb}{0.6, 0, 0}
\definecolor{tempText}{rgb}{0.55, 0.10,0.67}
\definecolor{eprintLinks}{rgb}{0.4, 0.4, 0.4}

\definecolor{journalLinks}{rgb}{0.6, 0, 0}

\titleformat{\section}[block]{\color{black}\Large\bfseries\filcenter}{\thetitle.\;}{0em}{}
\titleformat{\subsection}[block]{\color{black}\large\bfseries\filcenter}{}{0em}{}

\newcommand{\MYhref}[3][redLinks]{\href{#2}{\color{#1}{#3}}}%

\usepackage{multirow}
\let\orig@Hy@EveryPageAnchor\Hy@EveryPageAnchor
\def\Hy@EveryPageAnchor{%
    \begingroup
    \hypersetup{pdfview=Fit}%
    \orig@Hy@EveryPageAnchor
    \endgroup
}

\makeatother

\usepackage{babel}
\begin{document}
\title{ Flavorgenesis}
\author{Andreas Ekstedt\thanks{andreas.ekstedt@ipnp.mff.cuni.cz}\,, Renato
M. Fonseca\thanks{fonseca@ipnp.mff.cuni.cz}\, and Michal Malinsk\'{y}\thanks{malinsky@ipnp.troja.mff.cuni.cz}\date{}}
\maketitle

\lyxaddress{\begin{center}
{\Large{}\vspace{-0.5cm}}Institute of Particle and Nuclear Physics\\
Faculty of Mathematics and Physics, Charles University,\\
V Hole\v{s}ovi\v{c}k\'{a}ch 2, 18000 Prague 8, Czech Republic\\
\par\end{center}}
\begin{abstract}
We present a model where all fermions are contained in a single irreducible
representation of an $\SU{19}$ gauge symmetry group. If there is only one
scalar field, Yukawa interactions are controlled by a single
number rather than by one or more $3\times3$ matrices of couplings.
The low-energy concept of flavor emerges entirely from the scalar-sector
parameters; more specifically, entries of the
Standard Model Yukawa matrices are controlled by several vacuum
expectation values.
\end{abstract}

\section{Introduction}

The three-family fermion structure of the Standard Model is inherited by SU(5)  \cite{Georgi:1974sy} and
	SO(10) \cite{Georgi:1974my,Fritzsch:1974nn} grand unified theories. Therefore these models cannot explain the intricate
	pattern of masses and mixing angles observed at low energies, much less why there are three
	families in the first place.
	Yet if all fermions are part of a single irreducible representation $\Psi$ of a gauge group,
	then, given an appropriate scalar field $\Phi$, the Yukawa interactions

\begin{equation}
{\cal L}_Y=y\Psi\Psi\Phi
\end{equation}
would be controlled by a single number $y$. If feasible, such a setup would shed light on the origin of flavor.

But a viable model of this type also needs to reproduce numerous low-energy flavor observables (fermion masses and mixing parameters), which at first glance seems difficult.
Yet this worry is unwarranted because low-energy flavor parameters are not functions of $y$ alone. Indeed, the scalar $\Phi$ contains several singlets $\left(\boldsymbol{1},\boldsymbol{1},0\right)$ 
and doublets $\left(\boldsymbol{1},\boldsymbol{2},\pm1/2\right)$ of  $\mathrm{G}_\mathrm{SM}\equiv \SU{3}_\mathrm{C}\times \SU{2}_\mathrm{L}\times \mathrm{U}(1)_\mathrm{Y}$, and the low-energy-Yukawa interactions depend on their vacuum expectation values (VEVs).

If Nature behaves in this way, flavor, superficially looking
like a property of fermions, is actually an emergent phenomenon
originating exclusively from the scalar potential. We are therefore describing a mechanism of flavorgenesis.

Unification of all Standard Model fermion families within one~\cite{Georgi:1979md,Wilczek:1981iz}, or several ~\cite{Frampton:1979tj,Frampton:1979fd,Kim:1980ci}, irreducible representation of a unified group has been considered before, but none of these proposals correctly accommodate three\te and only three\te families of light fermions.

Recently it has 
been pointed out \cite{Fonseca:2015aoa,Yamatsu:2018tnv} that 
the three-family chiral structure of the Standard Model fits adequately into the 171-dimensional representation
of $\SU{19}$,\footnote{We only consider 4-dimensional space-time. Family
	unification has also been discussed in higher dimensions. In particular, reference \cite{Yamatsu:2018tnv} discusses some features of $\SU{19}$ family
	unification in 6 dimensions. See also reference \cite{Reig:2017nrz}.} and that this gauge theory might be uniquely suited for
flavor unification. In this letter we present a grand-unified model with the aforementioned structure that is capable of reproducing low-energy data. Explicit relations between
entries of the Standard Model Yukawa matrices and the vacuum expectation
values of the fundamental theory are constructed.

\section{The 171-dimensional representation of $\SU{19}$}

The fermion content of grand-unified theories is severely restricted by the requirement that only the Standard Model fermions are light.\footnote{Since they do no interact with the Standard Model gauge bosons, right-handed neutrinos,
	if they exist, might also be light.}
If there are no confining interactions \cite{Wilczek:1981iz}, additional fermions must be vector-like under $\mathrm{G}_\mathrm{SM}$ in order to obtain a large mass prior to electroweak symmetry breaking. That is, any new fermion $\mathrm{X}$
must be matched with an $\mathrm{X}^c$ having the same chirality and opposite quantum numbers.
Using this requirement, a systematic
study \cite{Fonseca:2015aoa} reveals that it is difficult
to unify all fermions in a single irreducible representation. Only one non-trivial case is known:\footnote{The Standard Model contains 45 fermions (distributed over 3 families),
	so it is trivially possible to embed these fields in the fundamental
	representation of the $\SU{45}$ group, or $\SU{45+n}$ if we account
	for $n$ right-handed neutrinos.} unification of all fermions in the $171$-dimensional
irreducible representation of $\SU{19}$. This uniqueness justifies a serious study of such a gauge theory.

Under $\SO{10}$,
a subgroup of $\SU{19}$, the $\boldsymbol{171}$ transforms as
\begin{equation}
\label{decomp171}
\underbrace{\boldsymbol{171}}_{\SU{19}}\rightarrow\underbrace{3\times\boldsymbol{16}+\boldsymbol{120}+3\times\boldsymbol{1}}_{\SO{10}}\,.
\end{equation}
The three $\mathbf{16}$s are know to contain all Standard Model fermions plus three right-handed neutrinos. This accounts for 48 out of the 171 components in equation \eqref{decomp171}.
 The remaining degrees of freedom transform as
real $\SO{10}$ representations ($\boldsymbol{120}$ or singlets) and are, therefore, vector-like under $\mathrm{G}_\mathrm{SM}$. All in all, 
only three Standard Model families of fermions are chiral. 
Yet this is only a counting argument; it
does not imply that all the light fermions are fully contained in the
three $\boldsymbol{16}$s. In fact, the $\boldsymbol{120}$ contains
two copies of the vector fermions $\left(\mathrm{d}^{c},\mathrm{d}\right)$
and $\left(\mathrm{L},\mathrm{L}^c\right)$, together with one copy of $\left(\mathrm{Q},\mathrm{Q}^c\right)$,
$\left(\mathrm{u}^{c},\mathrm{u}\right)$ and $\left(\mathrm{e}^{c},\mathrm{e}\right)$;
these states can mix with the ones in the three $\bf 16$s in equation (\ref{decomp171}).

Therefore, the relevant low-energy components are
\begin{equation}
\label{eq:3}
\underbrace{\boldsymbol{171}}_{\SU{19}}\rightarrow\underbrace{4\mathrm{Q}+4\mathrm{u}^{c}+5\mathrm{d}^{c}+5\mathrm{L}+4\mathrm{e}^{c}+\mathrm{Q}^c+\mathrm{u}+2\mathrm{d}+2\mathrm{L}^c+\mathrm{e}+\textrm{(more vector fermions)}}_{\SU{3}_\mathrm{C}\times \SU{2}_\mathrm{L}\times \mathrm{U}(1)_\mathrm{Y}}\,.
\end{equation}
From this discussion we conclude that, regardless of other details, this $\SU{19}$ fermion automatically reproduces the number of light fermions in the Standard Model.

It is worth noting that $\SU{19}$ contains a large number of inequivalent $\mathrm{G}_\mathrm{SM}$ subgroups,\footnote{There are thousands of such subgroups~\cite{Fonseca:2015aoa}.} which should not be surprising given the sizeable difference between the rank of the two groups. Yet we are only interested in a particular embedding, which can be understood as follows. One can embed $\mathrm{G}_\mathrm{SM}$ in $\SO{10}$
such that the spinor representation decomposes into a family
of SM fermions plus a right-handed neutrino, as usual. Since $\SO{10}$ is a special maximal subgroup of $\SU{16}$~\cite{Dynkin:1957um}, the fundamental representation of $\SU{16}$
is also irreducible under $\SO{10}$.\footnote{There are also models embedding a single Standard Model family, with or without right-handed neutrinos, into the fundamental representations of $\SU{15}$ or $\SU{16}$~\cite{Adler:1989nn,PhysRevD.25.3012,Pal:1990xw,PhysRevLett.64.619}.} Furthermore, the special unitary
group $\SU{19}$ contains $\SU{16} \times \SU{3}_\mathrm{F}\times \mathrm{U}(1)$
as a maximal subgroup~\cite{Yamatsu:2018tnv}. We may therefore write the following symmetry breaking chain

\begin{equation}
\underbrace{\boldsymbol{19}}_{\SU{19}}\rightarrow\underbrace{\left(\boldsymbol{16},\boldsymbol{1}\right)+\left(\boldsymbol{1},\boldsymbol{3}\right)}_{\SO{10}\times \SU{3}_\mathrm{F}}\rightarrow\underbrace{\mathbb{Q}+\mathbb{u^{c}}+\mathbb{d^{c}}+\mathbb{L}+\mathbb{e^{c}}+\mathbb{N^{c}}+\mathbb{N}_{k}^{\mathbb{c}}}_{\mathrm{G}_\mathrm{SM}\times \SU{3}_\mathrm{F}}\,~\hspace*{0.2cm}(k=1,2,3)\label{eq:4}
\end{equation}

The $\SU{3}_\mathrm{F}$ group is important\te even though
it must be broken\te  because it commutes with $\mathrm{G}_\mathrm{SM}$ and, as such, plays the role of a flavor group. For this reason it is worth keeping track of the transformation properties of the various fields under $\SU{3}_\mathrm{F}$.
The first six terms on the right hand side of branching rule~(\ref{eq:4}) stand for
the well known quantum numbers of the SM fermions plus
sterile neutrinos, which are all singlets under $\SU{3}_\mathrm{F}$. On the other hand,
the irreducible representation $\mathbb{N}_{k}^{\mathbb{c}}$ is invariant
under ${\rm G}_{\rm SM}$, but unlike $\mathbb{N^{c}}$
it is a triplet of $\SU{3}_\mathrm{F}$.

However, there is an even bigger $\SU{4}_\mathrm{F}$ that contains
$\SU{3}_\mathrm{F}$ and commutes with $\mathrm{G}_\mathrm{SM}$. Indeed, $\SU{5}\times \SU{4}_\mathrm{F}$ is
also a subgroup of  $\SU{19}$, under which
\begin{equation}
\underbrace{\boldsymbol{19}}_{\SU{19}}\rightarrow\underbrace{\left(\overline{\boldsymbol{5}},\boldsymbol{1}\right)+\left(\boldsymbol{10},\boldsymbol{1}\right)+\left(\boldsymbol{1}\boldsymbol{,4}\right)}_{\SU{5}\times \SU{4}_\mathrm{F}}\rightarrow\underbrace{\mathbb{Q}+\mathbb{u^{c}}+\mathbb{d^{c}}+\mathbb{L}+\mathbb{e^{c}}+\mathbb{N}_{i}^{\mathbb{c}}}_{\mathrm{G}_\mathrm{SM}\times \SU{4}_\mathrm{F}}\ \hspace*{0.2cm}(i=1,..,4),\label{eq:SU(5)xSU(4)-branching}
\end{equation}
where $\mathbb{N}_{i}^{\mathbb{c}}$ is a quadruplet under $\SU{4}_\mathrm{F}$.  We use lower Latin
indices for it, reserving upper Latin indices for an anti-quadruplet.
This $\SU{4}_\mathrm{F}$ is noteworthy for being the biggest simple subgroup of $\SU{19}$ which commutes with $\mathrm{G}_\mathrm{SM}\subset \SU{19}$.
There is a reason why we have been writing $\mathbb{Q}$, $\mathbb{u^{c}}$,
... (group representations) instead of $\mathrm{Q}$, $\mathrm{u}^{c}$,
... (actual fields): while the $\SU{19}$ model does not contain
a 19-dimensional representation, the $\mathbf{19}$'s decomposition  is crucial to the identification
of the various fermion and scalar components used.

The 171-dimensional fermion representation corresponds to the
anti-symmetric product of two fundamental $\SU{19}$ representations.
We can therefore view all fermions of equation~\eqref{eq:3} as belonging to a $19\times19$ anti-symmetric
matrix with the following block form

\renewcommand{\kbldelim}{(}
\renewcommand{\kbrdelim}{)}

\begin{equation}
\boldsymbol{171}= \kbordermatrix{ & \mathbb{d^{c}} & \mathbb{L} & \mathbb{Q} & \mathbb{u^{c}} & \mathbb{e^{c}} & \mathbb{N}_{i}^{\mathbb{c}}\\ \mathbb{d^{c}} & \mathrm{u} & \mathrm{Q}^{c} & \mathrm{L}_{1}^{c} & \mathrm{d}_{2} & \cdot & \mathrm{d}_{i}^{c}\\ \mathbb{L} & \times & \mathrm{e} & \mathrm{d}_{1} & \cdot & \mathrm{L}_{2}^{c} & \mathrm{L}_{i}\\ \mathbb{Q} & \times & \times & \cdot & \mathrm{L}_{5} & \cdot & \mathrm{Q}_{i}\\ \mathbb{u^{c}} & \times & \times & \times & \cdot & \mathrm{d}_{5}^{c} & \mathrm{u}_{i}^{c}\\ \mathbb{e^{c}} & \times & \times & \times & \times & \cdot & \mathrm{e}_{i}^{c}\\ \mathbb{N}_{j}^{\mathbb{c}} & \times & \times & \times & \times & \times & \mathrm{N}_{ij}^{c}
}\, \frac{1}{\sqrt{2}}.\label{eq:171-blocks}
\end{equation}For example, the block $\boldsymbol{171}_{\mathbb{L}\mathbb{N}_{i}^{\mathbb{c}}}$
has the ${\rm G}_{\rm SM}$ quantum numbers $\mathbb{L}\times\mathbb{N}_{i}^{\mathbb{c}}=\left(\boldsymbol{1},\boldsymbol{2},-1/2\right)$,
hence the name $\mathrm{L}_{i}$.
It is also a quadruplet of $\SU{4}_\mathrm{F}$, unlike
 $\mathrm{L}_{5}\equiv\boldsymbol{171}_{\mathbb{Q}\mathbb{u}^{\mathbb{c}}}$
which is a singlet of the flavor group. In total there are 4+1 $\mathrm{L}$ fields
in the $\boldsymbol{171}$. Equation \eqref{eq:171-blocks} encodes this information for all field components relevant at low energies.\footnote{Some blocks in equation~(\ref{eq:171-blocks}) contain fields beyond the ones shown here. For
example $\boldsymbol{171}_{\mathbb{Q}\mathbb{u}^{\mathbb{c}}}$ contains
both the $\left(\boldsymbol{1},\boldsymbol{2},+1/2\right)$ state, which
we called ${\rm L}_{5}$, and also $\left(\boldsymbol{8},\boldsymbol{2},+1/2\right)$ which is vector-like and, thus, irrelevant at low energies.} Also note that the right-handed neutrinos
$\mathrm{N}_{ij}^{c}\equiv\boldsymbol{171}_{\mathbb{N}_{i}^{\mathbb{c}}\mathbb{N}_{j}^{\mathbb{c}}}$ correspond to a sextet of ${\rm SU}(4)_{\rm F}$ (with $ij$ 
anti-symmetrized).

\section{The genesis of flavor}
There are two possible ways to couple two 171-dimensional representations to a single scalar field $\Phi$,
\begin{equation}
{\cal L}_Y=y\boldsymbol{171}_{ab}\boldsymbol{171}_{cd}\Phi^{abcd},\label{eq:fundamental-yukawa}
\end{equation}
corresponding to either a completely antisymmetric $\overline{\boldsymbol{3876}}$ or a mixed-symmetric $\overline{\boldsymbol{10830}}$  irreducible representation of $\SU{19}$. In what follows we consider the former case, ignoring for the moment the scalar potential and
possible issues with breaking the $\SU{19}$ gauge symmetry group down to~$\mathrm{G}_\mathrm{SM}$.

To shed light on flavor one must establish a connection
between the Standard Model Yukawa couplings and the parameters of the $\SU{19}$-phase Lagrangian.
For that sake we only need to consider those components of
$\Phi$ which can influence the light fermion spectrum: those that transform as 
$\left(\boldsymbol{1},\boldsymbol{1},0\right)$,
$\left(\boldsymbol{1},\boldsymbol{2},+1/2\right)$, and $\left(\boldsymbol{1},\boldsymbol{2},-1/2\right)$
under $\mathrm{G}_\mathrm{SM}$. We shall denote these types of fields by generic symbols $\mathrm{S}$, $\mathrm{H}$ and
$\widetilde{\mathrm{H}}$, respectively. 

There are 15 $\mathrm{S}$-type singlets, 11 $\mathrm{H}$'s, and 17 $\widetilde{\mathrm{H}}$'s
in $\Phi=\overline{\boldsymbol{3876}}$. Using the block notation introduced
above it is straightforward to identify their location; for example,
$\overline{\boldsymbol{3876}}^{\mathbb{Ld^{c}d^{c}u^{c}}}$ contains
an $\mathrm{H}$ doublet. All $\mathrm{S}$'s, $\mathrm{H}$'s, and $\tilde{\mathrm{H}}$'s are found in the blocks

\begin{align}
	\mathrm{S}:\; & \overbrace{\mathbb{d^{c}QQu^{c}}}^{\mathrm{S}_\mathrm{L}},\overbrace{\mathbb{LQu^{c}e^{c}}}^{\mathrm{S}_\mathrm{DL}},\overbrace{\mathbb{d^{c}u^{c}u^{c}e^{c}}}^{\mathrm{S}_\mathrm{D}},\overbrace{\mathbb{d^{c}d^{c}u^{c}}\mathbb{N}_{i}^{\mathbb{c}}}^{\mathrm{S}_\mathrm{UD}^{i}},\overbrace{\mathbb{d^{c}LQ}\mathbb{N}_{i}^{\mathbb{c}}}^{\mathrm{S}_\mathrm{QDL}^{i}},\overbrace{\mathbb{LLe^{c}}\mathbb{N}_{i}^{\mathbb{c}}}^{\mathrm{S}_{\mathrm{EL}}^{i}},\overbrace{\mathbb{N}_{i}^{\mathbb{c}}\mathbb{N}_{j}^{\mathbb{c}}\mathbb{N}_{k}^{\mathbb{c}}\mathbb{N}_{l}^{\mathbb{c}}}^{\mathrm{S}_\mathrm{N}}\,,\label{eq:S}
	\\
\mathrm{H}:\; & \mathbb{Ld^{c}d^{c}u^{c}},\mathbb{d^{c}LLQ},\overbrace{\mathbb{Qu^{c}}\mathbb{N}_{i}^{\mathbb{c}}\mathbb{N}_{j}^{\mathbb{c}}}^{\mathrm{H}_{\mathrm{QN}}^{ij}},\overbrace{\mathbb{L}\mathbb{N}_{j}^{\mathbb{c}}\mathbb{N}_{k}^{\mathbb{c}}\mathbb{N}_{l}^{\mathbb{c}}}^{\mathrm{H}_{\mathrm{N},i}}\,,\label{eq:H}\\
\widetilde{\mathrm{H}}:\; & \mathbb{d^{c}d^{c}d^{c}L},\overbrace{\left(\mathbb{Qu^{c}e^{c}}\mathbb{N}_{i}^{\mathbb{c}}\right)}^{\widetilde{\mathrm{H}}_{\mathrm{DE}}^{i}},\overbrace{\left(\mathbb{d^{c}Q}\mathbb{N}_{i}^{\mathbb{c}}\mathbb{N}_{j}^{\mathbb{c}}\right)}^{\widetilde{\mathrm{H}}_{\mathrm{D}}^{ij}},\overbrace{\left(\mathbb{Le^{c}}\mathbb{N}_{i}^{\mathbb{c}}\mathbb{N}_{j}^{\mathbb{c}}\right)}^{\widetilde{\mathrm{H}}_{\mathrm{E}}^{ij}}\,.\label{eq:Ht}
\end{align}\label{eq:ElectroWeakSinglets}
For later convenience, specific names were given to most of
these fields. Note that, just like fermions, some of these scalars
are singlets of the $\SU{4}_\mathrm{F}$ flavor group (for example $\mathrm{S}_\mathrm{L}$),
while others are anti-quadruplets with an upper index (such as $\widetilde{\mathrm{H}}_\mathrm{DE}^{i}$)
and others yet are anti-sextets ($\mathrm{H}_\mathrm{QN}^{ij}$ is one such case).
The only scalar which transforms as a quadruplet of $\SU{4}_{\rm F}$ is $\mathrm{H}_{{\rm N},i}\equiv\epsilon_{ijkl}\overline{\boldsymbol{3876}}^{\mathbb{L}\mathbb{N}_{j}^{\mathbb{c}}\mathbb{N}_{k}^{\mathbb{c}}\mathbb{N}_{l}^{\mathbb{c}}}$.

All the  fermion and scalar components that are important for the low-energy
phenomenology have been identified in equations (\ref{eq:171-blocks})
and (\ref{eq:S})--(\ref{eq:Ht}), but not all of them are light.
In particular, the VEVs of the electroweak singlets\te the $\mathrm{S}$'s\te determine which combinations of the $\mathrm{Q}$, $\mathrm{u}^{c}$,
$\mathrm{d}^{c}$, $\mathrm{L}$, and $\mathrm{e}^{c}$ fields become super-heavy. Furthermore, it is necessary to consider some fine tuning of scalar parameters, so that 
a linear combination of all the $\mathrm{H}$'s and $\widetilde{\mathrm{H}}^{*}$'s
forms the light Standard Model scalar doublet.

Expanding the $\SU{19}$-invariant Yukawa interaction in expression (\ref{eq:fundamental-yukawa}), we find 
\arraycolsep=1.5pt 
\begin{align}
& {\cal L}_Y\supset\frac{1}{3}\mathrm{Q}_{i}\mathrm{Q}^{c}\mathrm{S}_\mathrm{QDL}^{i}+\frac{\sqrt{2}}{3}\mathrm{u}_{i}^{c}\mathrm{u}\mathrm{S}_\mathrm{UD}^{i}+\sqrt{\frac{2}{3}}\mathrm{e}_{i}^{c}\mathrm{e}\mathrm{S}_\mathrm{EL}^{i}+\frac{1}{4}\sqrt{\frac{2}{3}}\epsilon_{ijkl}\mathrm{N}_{ij}^{c}\mathrm{N}_{kl}^{c}\mathrm{S}_\mathrm{N}\nonumber \\
 & +\frac{\sqrt{2}}{3}\left(\begin{array}{c}
\mathrm{d}_{i}^{c}\\
\mathrm{d}_{5}^{c}
\end{array}\right)^{T}\left(\begin{array}{cc}
-\mathrm{S}_\mathrm{QDL}^{i} & \sqrt{2}\mathrm{S}_\mathrm{UD}^{i}\\
-\mathrm{S}_\mathrm{DL} & \sqrt{2}\mathrm{S}_\mathrm{D}
\end{array}\right)\left(\begin{array}{c}
\mathrm{d}_{1}\\
\mathrm{d}_{2}
\end{array}\right)+\frac{1}{\sqrt{3}}\left(\begin{array}{c}
\mathrm{L}_{i}\\
\mathrm{L}_{5}
\end{array}\right)^{T}\left(\begin{array}{cc}
-\mathrm{S}_\mathrm{QDL}^{i} & \sqrt{2}\mathrm{S}_\mathrm{EL}^{i}\\
-\frac{2}{\sqrt{3}}\mathrm{S}_\mathrm{L} & \mathrm{S}_\mathrm{DL}
\end{array}\right)\left(\begin{array}{c}
\mathrm{L}_{1}^{c}\\
\mathrm{L}_{2}^{c}
\end{array}\right)\nonumber \\
 & -\frac{2}{3}\mathrm{Q}_{i}\mathrm{H}_\mathrm{QN}^{ij}\mathrm{u}_{j}^{c}+\frac{\sqrt{2}}{3}\left(\begin{array}{c}
 	\mathrm{d}_{i}^{c}\\
 	\mathrm{d}_{5}^{c}
 \end{array}\right)^T\left(\begin{array}{cc}
\sqrt{2}\widetilde{\mathrm{H}}_\mathrm{D}^{ij}\\  \widetilde{\mathrm{H}}_\mathrm{DE}^{j}\end{array}\right)\mathrm{Q}_{j}+\frac{1}{\sqrt{3}}\left(\begin{array}{c}
\mathrm{L}_{i}\\
\mathrm{L}_{5}
\end{array}\right)^{T}\left(\begin{array}{c}
-\epsilon_{ijkl}\mathrm{H}_{\mathrm{N},l}\\
\sqrt{2}\mathrm{H}_\mathrm{QN}^{jk}
\end{array}\right)\mathrm{N}_{jk}^{c}\nonumber \\
 & +\frac{\sqrt{2}}{3}\left(\begin{array}{c}
\mathrm{L}_{i}\\
\mathrm{L}_{5}
\end{array}\right)^{T}\left(\begin{array}{c}
-\sqrt{2}\widetilde{\mathrm{H}}_\mathrm{E}^{ij}\\
\widetilde{\mathrm{H}}_\mathrm{DE}^{j}
\end{array}\right)\begin{array}{c}
\mathrm{e}_{j}^{c}\end{array} \label{eq:main-result}
\end{align}
\arraycolsep=5.0pt plus other terms which are unimportant. Flavorgenesis is manifest in this expression: family-replication emerges from a fundamental theory which had none, and the flavor structure seen at low energies is a result of an alignment of VEVs.

As an example, consider the up-quarks. The four-component VEV $\mathrm{S}_\mathrm{QDL}^{i}$ defines
the heavy combination of the four $\mathrm{Q}$'s, while the VEV $\mathrm{S}_\mathrm{UD}^{i}$ plays an analogous role for
the $\mathrm{u}^{c}$'s. Therefore, these VEVs define $3\times4$ semi-unitary
matrices $\mathrm{U}_\mathrm{Q}$ and $\mathrm{U}_{\mathrm{u}^{c}}$ that project out the light fields.
As for the scalar doublets, their projection onto the Standard Model Higgs ($\mathrm{H}_\mathrm{X}\equiv\Lambda_\mathrm{X} \mathrm{H}_\mathrm{SM}+\text{heavy components}$, $\widetilde{\mathrm{H}}_\mathrm{Y}\equiv\Lambda_\mathrm{Y} \mathrm{H}^*_\mathrm{SM}+\text{heavy components}$ with $|\Lambda_\mathrm{X,Y}|\leq 1$) can be determined from the scalar potential.
 Note that in the corresponding relations for $\SU{4}_{\rm F}$ sextets like $\mathrm{H}_\mathrm{QN}^{ij}=\Lambda_\mathrm{QN}^{ij}\mathrm{H}_\mathrm{SM}+\cdots$ the proportionality coefficients 
$\Lambda_\mathrm{QN}$ form a $4\times4$ anti-symmetric matrix; these coefficients are in general complicated functions of the scalar potential parameters.

With all this at hand, the Standard Model up-quark Yukawa matrix $\mathrm{Y}_\mathrm{U}$ at the unification scale is computed readily:
\begin{equation}
\mathrm{Y}_\mathrm{U}=-\frac{2}{3}\mathrm{U}_\mathrm{Q}\Lambda_\mathrm{QN}\mathrm{U}_{\mathrm{u}^{c}}^{T}\,.
\end{equation}
A similar but somewhat more complicated analysis can be made for down
quarks, charged leptons and neutrinos.

As for the neutrinos, note that there are five different types of VEVs (those of $\mathrm{S}^i_\mathrm{\rm QDL}$, $\mathrm{S}^i_\mathrm{\rm EL}$, $\mathrm{S}_\mathrm{\rm L}$, $\mathrm{S}_\mathrm{\rm DL}$  and $\mathrm{S}_\mathrm{N}$) contributing to the light neutrino masses through a rather elaborate seesaw mechanism. One finds that if $\left\langle \mathrm{S}_{\mathrm{N}}\right\rangle$ is significantly smaller than the remaining VEVs\te assumed to trigger the $\SU{19}$ symmetry breaking\te then one of the light neutrinos is practically massless.

Besides this observation, we have checked that it is possible to reproduce any pattern of the Standard Model fermion masses and mixing angles, if the VEVs of the scalars  ($\mathrm{S}$, $\mathrm{H}$ and
$\widetilde{\mathrm{H}}$) are treated as free. This means that expression \eqref{eq:main-result} is not ruled out by data, but at the same time, none of the Standard Model flavor parameters can be predicted without looking carefully into the scalar potential. 

\section{Symmetry Breaking\label{sec:SymmetryBreaking}}
It is tempting to consider that the very same scalar representation $\Phi$ that interacts with fermions is also responsible for breaking $\SU{19}$ to $\mathrm{G}_\text{SM}$; if possible, this would be unlike, for example, $\SO{10}$ and $\mathrm{E}_6$ grand unified theories where multiple scalar irreducible representations are typically needed.
 And in our model the required group-rank reduction is even bigger\te from 18 to 4\te so it seems hard to do so with a single irreducible representation.
However, the current scalar is a four-index-antisymmetric tensor, for which many breaking patterns are possible ~\cite{Cummins:1985vg,Cummins:1984wt,Jetzer:1983ij,Basecq:1986yg,Hubsch:1984zi}.

Indeed, our analysis shows that $\Phi=\overline{\mathbf{3876}}$ has the right components to break $\SU{19}$ into $\mathrm{G}_\mathrm{SM}$, and the VEVs that achieve this breaking are solutions of the tadpole equations. Yet as far as we could tell, for all of these VEVs there are tachyonic scalars in the spectrum.
The rest of this section expands on these statements.
  
It is a direct consequence of the branching rule \eqref{eq:SU(5)xSU(4)-branching} that $\mathrm{F}\equiv\SU{4}_\mathrm{F}\times \mathrm{U}(1)^4$ is the biggest subgroup of $\SU{19}$ which commutes with $\mathrm{G}_\text{SM}$, so at the very least $\mathrm{F}$ must be completely broken. 
In order to break $\SU{4}_\mathrm{F}$, three scalars in the fundamental representation, with linearly independent VEVs, are needed~\cite{Li:1973mq}.
Moreover, on very general grounds, one needs a different VEV (which does not need to be charged under $\SU{4}_\mathrm{F}$) to break each of the remaining $\mathrm{U}(1)$s. This suggests that $\Phi$ should contain at least 3 S's that are quadruplets of the flavour group, and 4 S's that are singlets under it. It turns out that this corresponds exactly to the set of available  $\mathrm{G}_\mathrm{SM}$ preserving directions in $\Phi$, see expression \eqref{eq:ElectroWeakSinglets}.

However, this counting is only indicative as, in some cases, the  symmetry-breaking power of the  ${\rm S}$'s may be further enhanced: arranging, for instance, the 3 quadruplet VEVs $\lra{\mathrm{S}^i_\mathrm{QDL}}$, $\lra{\mathrm{S}^i_\mathrm{UD}}$ and $\lra{\mathrm{S}^i_\mathrm{EL}}$ to be linearly independent but not perpendicular to each other ($\sum_i \lra{S^i_\mathrm{X}}\lra{S^i_\mathrm{Y}}\neq 0$), it is possible to break the $\SU{4}_\mathrm{F}$ flavor group completely along with two extra $\mathrm{U}(1)$'s. As a consequence, the VEVs of up to two of the $\SU{4}_\mathrm{F}$-invariant S's can even be null and, yet, the entire $\SU{4}_\mathrm{F}\times
 \mathrm{U}(1)^4$ factor together with any residual symmetry between $\SU{19}$ and $\mathrm{G}_\mathrm{SM}$ (such as $\SU{5}$ in equation~(\ref{eq:SU(5)xSU(4)-branching})) may still be broken appropriately.

As an example consider the following vacuum direction
\begin{align}\label{VEVdirection}
&\lra{ \mathrm{S}^i_\mathrm{QDL}}=(v^1_\mathrm{QDL},v^2_\mathrm{QDL},0,0),~\lra{\mathrm{S}^i_\mathrm{UD}}=(v_\mathrm{UD}^1,v_\mathrm{UD}^2,0,0),~\lra{\mathrm{S}^i_\mathrm{EL}}=(0,0,v_\mathrm{EL},0)\,,
\\& \lra{\mathrm{S}_\mathrm{DL}}=0,~\lra{\mathrm{S}_\mathrm{D}}=v_\mathrm{D},~\lra{\mathrm{S}_\mathrm{L}}=v_\mathrm{L},~\lra{\mathrm{S}_\mathrm{N}}=v_\mathrm{N}.\nonumber
\end{align}
It can be shown that this configuration 
breaks the $\SU{19}$ gauge symmetry down to $\mathrm{G}_\mathrm{SM}$ and that it is also a solution of the tadpole equations for the renormalizable potential
\begin{align}
\mathrm{V}=-m^{2}\left(\Phi\Phi^{*}\right)+\lambda_{1}\left(\Phi\Phi^{*}\right)^{2}+\lambda_{2}\left(\Phi\Phi^{*}\right)_{b}^{a}\left(\Phi\Phi^{*}\right)_{a}^{b}+\lambda_{3}\left(\Phi\Phi^{*}\right)_{ab}^{cd}\left(\Phi\Phi^{*}\right)_{cd}^{ab}
\,.\label{eq:ScalarPotential}
\end{align}
Here we used the abbreviations $\left(\Phi\Phi^{*}\right)\equiv\Phi^{abcd}\Phi_{abcd}^{*}$ , $\left(\Phi\Phi^{*}\right)_{y}^{x}\equiv\Phi^{xbcd}\Phi_{ybcd}^{*}$
and $\left(\Phi\Phi^{*}\right)_{yy^{\prime}}^{xx^{\prime}}\equiv\Phi^{xx^{\prime}cd}\Phi_{yy^{\prime}cd}^{*}$.

Note that this potential has a $\mathrm{U}(19)$ symmetry~\cite{Cummins:1985vg}, which includes the gauge group and a global $\mathrm{U}(1)$ associated to a rephasing of the entire $\Phi$ scalar.

For the VEV direction~(\ref{VEVdirection}) we have obtained the correct number of would-be Goldstone bosons corresponding to the $\SU{19}/\mathrm{G}_\mathrm{SM}$ coset.\footnote{The global $\mathrm{U}(1)$ rephasing symmetry is also broken, so one would think that there is a truly massless Goldstone boson. However, this is not the case because a mixture of this rephasing $\mathrm{U}(1)$ and the gauge symmetry generators is unbroken.} However, for all such solutions there are always some tachyonic scalars in the spectrum. This issue can possibly be overcome if we admit all S's to take non-zero VEVs. But such a case is arduous to analyse so we leave this question open. It could also be worth considering radiative corrections~\cite{Bertolini:2009es}.

\section{Final thoughts}
In this paper we have presented a potentially realistic unified model where all fermions are contained in a single irreducible representation of the $\SU{19}$ gauge group; thus, a single number controls all Yukawa interactions at the fundamental level. The Standard Model family replication is an emergent phenomenon only manifest at low energies. We have shown explicitly that the Standard Model fermion masses and mixing parameters can depend exclusively on the VEVs of a single scalar field; in this sense the concept of flavor originates entirely from the scalar sector.

Furthermore, we found that the model is compatible with the low-energy data if all VEVs are taken as free parameters. Such a consistency check does not provide any further insight into the fermion masses and mixing parameters, but this could change with a dedicated analysis of the scalar potential.

Remarkably, the same scalar which governs all the Yukawa interactions is capable of breaking the original gauge symmetry all the way down to $\mathrm{G}_\mathrm{SM}$. An initial study of the tree-level potential suggests that tachyonic scalars are a problem, but a more thorough analysis might still reveal viable points in the parameter space. If that is not the case, the situation can be amended by adding another scalar or perhaps by considering radiative corrections.

Due to its fermion content, the model has a gauge anomaly. This shortcoming may be resolved in various ways without affecting its main features (but perhaps making the whole setup less elegant); for example, only part of the symmetry group might be gauged, or further fermionic representations can be added within a confining sector. 

\section*{Acknowledgments}
We acknowledge the financial support from the Grant Agency of the
Czech Republic (GA\v{C}R) through contract number 20-17490S and from
the Charles University Research Center UNCE/SCI/013.


\begin{thebibliography}{10}
	\providecommand{\url}[1]{\texttt{#1}}
	\providecommand{\urlprefix}{URL }
	\providecommand{\eprint}[2][]{\url{#2}}
	
	\bibitem{Georgi:1974sy}
	H.~Georgi and S.~Glashow, \emph{{Unity of all elementary particle forces}},
	\MYhref[journalLinks]{http://dx.doi.org/10.1103/PhysRevLett.32.438}{Phys.
		Rev. Lett.
	}\MYhref[journalLinks]{http://dx.doi.org/10.1103/PhysRevLett.32.438}{\textbf{32}
		(1974) 438--441}.
	
	\bibitem{Georgi:1974my}
	H.~Georgi, \emph{{The state of the art: gauge theories}},
	\MYhref[journalLinks]{http://dx.doi.org/10.1063/1.2947450}{AIP Conf. Proc.
	}\MYhref[journalLinks]{http://dx.doi.org/10.1063/1.2947450}{\textbf{23}
		(1975) 575--582}.
	
	\bibitem{Fritzsch:1974nn}
	H.~Fritzsch and P.~Minkowski, \emph{{Unified interactions of leptons and
			hadrons}},
	\MYhref[journalLinks]{http://dx.doi.org/10.1016/0003-4916(75)90211-0}{Annals
		Phys.
	}\MYhref[journalLinks]{http://dx.doi.org/10.1016/0003-4916(75)90211-0}{\textbf{93}
		(1975) 193--266}.
	
	\bibitem{Georgi:1979md}
	H.~Georgi, \emph{{Towards a grand unified theory of flavor}},
	\MYhref[journalLinks]{http://dx.doi.org/10.1016/0550-3213(79)90497-8}{Nucl.
		Phys. B
	}\MYhref[journalLinks]{http://dx.doi.org/10.1016/0550-3213(79)90497-8}{\textbf{156}
		(1979) 126--134}.
	
	\bibitem{Wilczek:1981iz}
	F.~Wilczek and A.~Zee, \emph{{Families from spinors}},
	\MYhref[journalLinks]{http://dx.doi.org/10.1103/PhysRevD.25.553}{Phys. Rev. D
	}\MYhref[journalLinks]{http://dx.doi.org/10.1103/PhysRevD.25.553}{\textbf{25}
		(1982) 553}.
	
	\bibitem{Frampton:1979tj}
	P.~Frampton, \emph{{Unification of flavor}},
	\MYhref[journalLinks]{http://dx.doi.org/10.1016/0370-2693(80)90140-9}{Phys.
		Lett. B
	}\MYhref[journalLinks]{http://dx.doi.org/10.1016/0370-2693(80)90140-9}{\textbf{89}
		(1980) 352--354}.
	
	\bibitem{Frampton:1979fd}
	P.~Frampton and S.~Nandi, \emph{{SU(9) grand unification of flavor with three
			generations}},
	\MYhref[journalLinks]{http://dx.doi.org/10.1103/PhysRevLett.43.1460}{Phys.
		Rev. Lett.
	}\MYhref[journalLinks]{http://dx.doi.org/10.1103/PhysRevLett.43.1460}{\textbf{43}
		(1979) 1460}.
	
	\bibitem{Kim:1980ci}
	J.~E. Kim, \emph{{A model of flavor unity}},
	\MYhref[journalLinks]{http://dx.doi.org/10.1103/PhysRevLett.45.1916}{Phys.
		Rev. Lett.
	}\MYhref[journalLinks]{http://dx.doi.org/10.1103/PhysRevLett.45.1916}{\textbf{45}
		(1980) 1916}.
	
	\bibitem{Fonseca:2015aoa}
	R.~M. Fonseca, \emph{{On the chirality of the SM and the fermion content of
			GUTs}},
	\MYhref[journalLinks]{http://dx.doi.org/10.1016/j.nuclphysb.2015.06.012}{Nucl.
		Phys. B
	}\MYhref[journalLinks]{http://dx.doi.org/10.1016/j.nuclphysb.2015.06.012}{\textbf{897}
		(2015) 757--780},
	\MYhref[eprintLinks]{http://arxiv.org/abs/1504.03695}{{\ttfamily
			arXiv:1504.03695 [hep-ph]}}.
	
	\bibitem{Yamatsu:2018tnv}
	N.~Yamatsu, \emph{{Family unification in special grand unification}},
	\MYhref[journalLinks]{http://dx.doi.org/10.1093/ptep/pty100}{PTEP
	}\MYhref[journalLinks]{http://dx.doi.org/10.1093/ptep/pty100}{\textbf{2018}
		(2018) 9 091B01},
	\MYhref[eprintLinks]{http://arxiv.org/abs/1807.10855}{{\ttfamily
			arXiv:1807.10855 [hep-ph]}}.
	
	\bibitem{Reig:2017nrz}
	M.~Reig, J.~W.~F. Valle, C.~Vaquera-Araujo and F.~Wilczek, \emph{{A model of
			comprehensive unification}},
	\MYhref[journalLinks]{http://dx.doi.org/10.1016/j.physletb.2017.10.038}{Phys.
		Lett. B
	}\MYhref[journalLinks]{http://dx.doi.org/10.1016/j.physletb.2017.10.038}{\textbf{774}
		(2017) 667--670},
	\MYhref[eprintLinks]{http://arxiv.org/abs/1706.03116}{{\ttfamily
			arXiv:1706.03116 [hep-ph]}}.
	
	\bibitem{Dynkin:1957um}
	E.~Dynkin, \emph{{Semisimple subalgebras of semisimple Lie algebras}}, Trans.
	Am. Math. Soc. \textbf{6} (1957) 111--244.
	
	\bibitem{Adler:1989nn}
	S.~L. Adler, \emph{{A new electroweak and strong interaction unification
			scheme}},
	\MYhref[journalLinks]{http://dx.doi.org/10.1016/0370-2693(89)91025-3}{Phys.
		Lett. B
	}\MYhref[journalLinks]{http://dx.doi.org/10.1016/0370-2693(89)91025-3}{\textbf{225}
		(1989) 143}, [Erratum: Phys.Lett.B 228, 560 (1989)].
	
	\bibitem{PhysRevD.25.3012}
	R.~N. Mohapatra and M.~Popovi\ifmmode~\acute{c}\else \'{c}\fi{}, \emph{Maximal
		grand unification, gauge hierarchies, and baryon nonconservation},
	\MYhref[journalLinks]{http://dx.doi.org/10.1103/PhysRevD.25.3012}{Phys. Rev.
		D
	}\MYhref[journalLinks]{http://dx.doi.org/10.1103/PhysRevD.25.3012}{\textbf{25}
		(1982) 3012--3026}.
	
	\bibitem{Pal:1990xw}
	P.~B. Pal, \emph{{Mass scales and symmetry breaking in SU(15) grand
			unification}},
	\MYhref[journalLinks]{http://dx.doi.org/10.1103/PhysRevD.43.236}{Phys. Rev. D
	}\MYhref[journalLinks]{http://dx.doi.org/10.1103/PhysRevD.43.236}{\textbf{43}
		(1991) 236--240}.
	
	\bibitem{PhysRevLett.64.619}
	P.~H. Frampton and B.-H. Lee, \emph{S{U}(15) grand unification},
	\MYhref[journalLinks]{http://dx.doi.org/10.1103/PhysRevLett.64.619}{Phys.
		Rev. Lett.
	}\MYhref[journalLinks]{http://dx.doi.org/10.1103/PhysRevLett.64.619}{\textbf{64}
		(1990) 619--621}.
	
	\bibitem{Cummins:1985vg}
	C.~Cummins and R.~King, \emph{{Absolute minima of the Higgs potential for the
			75 of SU(5)}},
	\MYhref[journalLinks]{http://dx.doi.org/10.1088/0305-4470/19/2/013}{J. Phys.
		A
	}\MYhref[journalLinks]{http://dx.doi.org/10.1088/0305-4470/19/2/013}{\textbf{19}
		(1986) 161}.
	
	\bibitem{Cummins:1984wt}
	C.~Cummins and R.~King, \emph{{Symmetry breaking patterns for third rank
			totally antisymmetric tensor reprensentations of unitary groups}},
	\MYhref[journalLinks]{http://dx.doi.org/10.1088/0305-4470/17/12/001}{J. Phys.
		A
	}\MYhref[journalLinks]{http://dx.doi.org/10.1088/0305-4470/17/12/001}{\textbf{17}
		(1984) L627--L633}.
	
	\bibitem{Jetzer:1983ij}
	P.~Jetzer, J.~Gerard and D.~Wyler, \emph{{Possible symmetry breaking patterns
			with totally symmetric and antisymmetric representations}},
	\MYhref[journalLinks]{http://dx.doi.org/10.1016/0550-3213(84)90206-2}{Nucl.
		Phys. B
	}\MYhref[journalLinks]{http://dx.doi.org/10.1016/0550-3213(84)90206-2}{\textbf{241}
		(1984) 204--220}.
	
	\bibitem{Basecq:1986yg}
	J.~Basecq, S.~Meljanac and D.~Pottinger, \emph{{Stable absolute minima of Higgs
			potentials with high rank representations}},
	\MYhref[journalLinks]{http://dx.doi.org/10.1016/0550-3213(87)90643-2}{Nucl.
		Phys. B
	}\MYhref[journalLinks]{http://dx.doi.org/10.1016/0550-3213(87)90643-2}{\textbf{292}
		(1987) 222--236}.
	
	\bibitem{Hubsch:1984zi}
	T.~H{\"{u}}bsch, S.~Meljanac and S.~Pallua, \emph{{Symmetry breaking of SU(n)
			gauge theories to maximal regular subgroups and fourth rank tensors}},
	\MYhref[journalLinks]{http://dx.doi.org/10.1103/PhysRevD.31.352}{Phys. Rev. D
	}\MYhref[journalLinks]{http://dx.doi.org/10.1103/PhysRevD.31.352}{\textbf{31}
		(1985) 352}.
	
	\bibitem{Li:1973mq}
	L.-F. Li, \emph{{Group theory of the spontaneously broken gauge symmetries}},
	\MYhref[journalLinks]{http://dx.doi.org/10.1103/PhysRevD.9.1723}{Phys. Rev. D
	}\MYhref[journalLinks]{http://dx.doi.org/10.1103/PhysRevD.9.1723}{\textbf{9}
		(1974) 1723--1739}.
	
	\bibitem{Bertolini:2009es}
	S.~Bertolini, L.~Di~Luzio and M.~Malinsk{\'{y}}, \emph{{On the vacuum of the
			minimal nonsupersymmetric SO(10) unification}},
	\MYhref[journalLinks]{http://dx.doi.org/10.1103/PhysRevD.81.035015}{Phys.
		Rev.
	}\MYhref[journalLinks]{http://dx.doi.org/10.1103/PhysRevD.81.035015}{\textbf{D81}
		(2010) 035015}, \MYhref[eprintLinks]{http://arxiv.org/abs/arXiv:0912.1796
		[hep-ph]}{{\ttfamily arXiv:0912.1796 [hep-ph] [hep-ph]}}.
	
\end{thebibliography}
\end{document}